\documentclass[aps,pre,subeqn,nofootinbib,notitlepage,amsmath,amsfonts]{revtex4-1}
\usepackage{graphicx, xcolor}
\usepackage{longtable}
\usepackage{subfigure}
\usepackage{euscript}
\usepackage[hyperfootnotes=false]{hyperref}

\newcommand{\eqr}[1]{Eq.~\eqref{#1}}

\newcommand{\secr}[1]{Sec.~[\ref{#1}]}
\newcommand{\figr}[1]{Fig.~[\ref{#1}]}
\newcommand{\figsr}[2]{Fig.~[\ref{#1}-\ref{#2}]}

\newcommand{\tblr}[1]{Table~[\ref{#1}]}
\newcommand{\tblsr}[2]{Tables~[\ref{#1}-\ref{#2}]}
\newcommand{\appr}[1]{Appendix~[\ref{#1}]}

\newcommand{\velocity}{{\rm{v}}}

\newcommand{\spd}{\!\cdot\!}

\newcommand{\BigVarea}{\vphantom{\Big(\Big)}}
\newcommand{\bigVarea}{\vphantom{\big(\big)}}

\newcommand{\xs}{x^{\scriptstyle s}}

\newcommand{\xgsb}{x^{g,s}}
\newcommand{\xlsb}{x^{\ell,s}}
\newcommand{\xg}{x^{g}}
\newcommand{\xl}{x^{\ell}}

\newcommand{\profilescale}{0.4}

\newcommand{\wh}{\widetilde{h}}
\newcommand{\wmu}{\widetilde{\mu}}

\makeatletter%
\renewcommand*\env@matrix[1][c]{\hskip -\arraycolsep%
  \let\@ifnextchar\new@ifnextchar%
  \array{*\c@MaxMatrixCols #1}}%
\makeatother%

\graphicspath{{Figures/}}

\begin{document}
\title{Resistances for heat and mass transfer through a liquid-vapor interface in a binary mixture.}
\author{K.~S.~Glavatskiy$^{1,2}$}
\author{D.~Bedeaux$^{1,2}$}
\affiliation
{%
$^{1}$Department of Chemistry, Norwegian University of Science and Technology, NO 7491 Trondheim, Norway.\\
$^{2}$Department of Process and Energy, Technical University of Delft, Leeghwaterstr 44, 2628 CA Delft, The Netherlands.
}%
\date\today
\begin{abstract}
In this paper we calculate the interfacial resistances to heat and mass transfer through a liquid-vapor interface in a binary mixture. We use two methods, the
direct calculation from the actual non-equilibrium solution and integral relations, derived earlier. We verify, that integral relations, being a relatively faster
and cheaper method, indeed gives the same results as the direct processing of a non-equilibrium solution. Furthermore we compare the absolute values of the
interfacial resistances with the ones obtained from kinetic theory. Matching the diagonal resistances for the binary mixture we find that kinetic theory
underestimates the cross coefficients. The heat of transfer is as a consequence correspondingly larger.
\end{abstract}
\maketitle
%
%
\numberwithin{equation}{section}
\section{Introduction}

A number of different methods have been used to obtain the surface transfer coefficients for one-component systems: experiments \cite{Fang1999a, Bedeaux1999b,
Mills2002, James2006}, molecular dynamic simulations \cite{Rosjorde2000, Rosjorde2001, Kjelstrup2002, Simon2004, jialin/longrange} , kinetic theory \cite{Pao1971a,
Sone1973, Cipolla1974, Bedeaux1990}. In a paper coauthored by one of us \cite{bedeaux/vdW/III} the interfacial transfer coefficients were calculated with the square
gradient theory for a one-component system, and compared to the data in the above references. Even for one-component systems the database of interfacial transfer
coefficients is poor and these data are pretty scattered. The situation is even worse for mixtures. There are only few experiments available \cite{James2006,
Mills2002} at a very restrictive range of conditions, i.e. for instance, at infinite dilution. No molecular dynamic simulations are available yet. The only source
of values of interfacial coefficients is kinetic theory \cite{Cipolla1974, Bedeaux1990}. This theory is most appropriate for short range potentials and low density
gases. There is evidence from molecular dynamic simulations for one-component systems for longer range potentials \cite{jialin/longrange} that the coupling transfer
resistivities for liquid-vapor interfaces of real fluids are substantially larger than those predicted by kinetic theory.

It is the aim of this article to determine the heat and mass transfer resistances of the interfacial region. The values of these transfer coefficients, or even
their order of magnitude, are extremely important for industrial processes which involve evaporation and/or condensation of mixtures. Among these processes is, for
instance, distillation, when one needs to separate components with different volatilities. As this involves evaporation and/or condensation repeatedly many times,
it is very important to know the exact effect of the surface. Some values of the interfacial transfer coefficients may favor transport of a component, while other
values may not. Of particular interest are the values of the cross coefficients, which contribute to reversible transport, and which are in most descriptions
neglected \cite{kjelstrupbedeaux/heterogeneous}.

We will verify that integral relations, derived in \cite{glav/tc1} give the same values of resistances, obtained directly from a non-equilibrium numerical solution.
The numerical solution is obtained using the non-equilibrium square gradient model \cite{glav/grad2}. It is desirable to compare our predictions with other methods,
in particular molecular simulations and experiments. Such data are not available yet, however, and we will therefore use the predictions of kinetic theory to
compare with.

In our approach we use the local resistivity profiles. The values of the local resistivities in the liquid and the vapor phases are chosen on the basis of
experimental values. In the interfacial region there are small peaks in these resistivities. The results of molecular dynamics simulations \cite{Simon2004} support
the existence of such peaks in the local resistivities in the interfacial region. The amplitudes, being the adjustable parameters, control the magnitude of these
peaks. The square gradient approach gives a natural tool to incorporate these peak in the theory. Possible values of these amplitudes are found by matching the
diagonal transfer coefficients to values predicted by kinetic theory. Using these amplitudes we find that the value of the cross resistivities is 1-2 orders of
magnitude higher then the one from kinetic theory. The results indicate that kinetic theory underestimates the interfacial transfer coefficients in real fluids. One
of them even has a different sign.

Consider a planar interface between a liquid and a vapor of a mixture through which there is evaporation or condensation. The mixture is in a box with gravity $g$
directed along the $x$-axis from left to right. The gas phase is therefore in the left part of the box and the liquid is in the right part. Due to evaporation or
condensation there exists a mass flux $J_{\xi_i}$ of component $i$, which is equal to the mass of component $i$ transferred through a unit surface area per unit of
time. Furthermore, there exists the total energy flux $J_{e}$, which is defined similarly. In stationary conditions these fluxes are constants (independent of $x$).

A surface can be described by Gibbs excess properties. We refer to \cite{glav/tc1} and \cite{glav/grad2} for an explanation how these quantities can be introduced
in non-equilibrium. Due to non-equilibrium conditions the temperature and the chemical potentials of the components are not the same in the liquid and in the gas
phases. Let $T^{\ell}$ and $T^{g}$ be the extrapolated temperatures of these phases at the surface. The exact position of the dividing surface is irrelevant for the
following analysis. Furthermore, let $\mu^{\ell}_{i}$ and $\mu^{g}_{i}$ similarly be the extrapolated chemical potentials of the $i$-th component at the surface.

The paper is organized as follows. In \secr{sec/EntropyProduction} we discuss the different forms of the excess entropy production of the interface and introduce
interfacial resistances. \secr{sec/KineticTheory} gives the overview of the expressions for these coefficients predicted from kinetic theory. We further build a
procedure to determine the actual values of these resistances directly from a non-equilibrium numerical solution in \secr{sec/NonEqSolution} and from integral
relations, which use only equilibrium profiles in \secr{sec/IntegralRelations}. We compare the predictions of all three methods in \secr{sec/Results} and discuss
the results in \secr{sec/Discussion}.

\section{Excess entropy production}\label{sec/EntropyProduction}

In a previous paper \cite{glav/tc1} we have obtained the following relation for the excess entropy production for the Gibbs surface in case of transport through the
interface\footnote{In \cite{glav/tc1} we have used the notation $\mathfrak{E}\left[{\sigma}_{s}\right]$ for the excess entropy production to distinguish it from the
local entropy production ${\sigma}_{s}$. Here we do not use the local entropy production and therefore will denote the excess entropy production by ${\sigma}_{s}$
to simplify the notation.}
\begin{equation}\label{eq/ExcessEntropy/02}
{\sigma}_{s} = J_{e}\left(\frac{1}{T^{\ell}} - \frac{1}{T^{g}}\right) -  \sum_{i=1}^{n}{J_{\xi_{i}}\left(\frac{{\wmu}_{i}^{\ell}}{T^{\ell}}-
\frac{{\wmu}_{i}^{g}}{T^{g}}\right)}
\end{equation}%
where ${\wmu}_{i} \equiv \mu_{i} + \velocity^{2}/2-g\xs$, with $\velocity$ the barycentric velocity and $\xs$ the position of the dividing surface. We introduce the
measurable heat flux $J_{q}'$ by
\begin{equation}\label{eq/ExcessEntropy/03}
J_{q}' = J_{e} - \sum_{i=1}^{n}{{\wh}_{i}J_{\xi_i}}
\end{equation}%
where ${\wh}_{i} \equiv h_{i} + \velocity^{2}/2-g\xs = {\wmu}_{i} + Ts_{i}$, with $s_{i}$ the partial entropy and $h_{i}$ the partial enthalpy. Using the measurable
heat flux on the vapor side, the excess entropy production can then be written as \cite{glav/tc1}
\begin{equation}\label{eq/ExcessEntropy/10a}
{\sigma}_{s} = J_{q}^{\,\prime,\,g} \left(\frac{1}{T^{\ell}}-\frac{1}{T^{g}}\right) - \sum_{i=1}^{n}{J_{\xi_{i}}\left[%
\left(\frac{{\wmu}_{i}^{\ell}}{T^{\ell }}-\frac{{\wmu}_{i}^{g}}{T^{g}}\right) -{\wh}_{i}^{g}\left( \frac{1}{T^{\ell }}- \frac{1}{T^{g}}\right) \right] }
\end{equation}
An alternative form of this expression is
\begin{equation}\label{eq/ExcessEntropy/05a}
{\sigma}_{s} = J_{q}^{\,\prime,\,g}\left(\frac{1}{T^{\ell}} - \frac{1}{T^{g}}\right) - \sum_{i=1}^{n}{J_{\xi_{i}}\frac{1}{T^{\ell}}\left(\tilde{\mu}_{i}^{\ell}-
\tilde{\mu}_{i}^{g} + s_{i}^{g}(T^{\ell}-T^{g})\right)}
\end{equation}
It is important to realize that \eqr{eq/ExcessEntropy/05a}, \eqr{eq/ExcessEntropy/10a} and \eqr{eq/ExcessEntropy/02} are exactly equivalent. It is common to do
these transformations neglecting third and higher order contributions in the deviation from equilibrium. Such approximations were not needed here. If one neglects
such higher order terms one may write \eqr{eq/ExcessEntropy/05a} in the form
\begin{equation}\label{eq/ExcessEntropy/06a}
{\sigma}_{s} = J_{q}^{\,\prime,\,g}\left(\frac{1}{T^{\ell }}-\frac{1}{T^{g}}\right) -
\sum_{i=1}^{n}{J_{\xi_{i}}\frac{1}{T^{\ell}}\left({\wmu}_{i}^{\ell}-{\wmu}_{i}^{g}(T^{\ell})\right)}
\end{equation}
This expression is convenient if one wants to write the chemical forces in terms of the natural logarithm of the partial pressure divided by the partial vapor
pressure of the liquid\footnote{These partial pressures are defined as the molar concentrations times the total pressure.}. We refer to
\cite{kjelstrupbedeaux/heterogeneous} for a discussion of this.

\eqr{eq/ExcessEntropy/06a} has the form of the entropy production for the surface used in \cite{kjelstrupbedeaux/heterogeneous}. It was obtained there using the
local equilibrium hypothesis, which we have proven to be valid in \cite{glav/grad2}. In \cite{glav/tc1} we have derived \eqr{eq/ExcessEntropy/05a} independently, by
calculating the excess of the continuous entropy production.

We now consider a binary mixture. The excess entropy production can be written as
\begin{equation}\label{eq/TwoComponent/00}
{\sigma}_{s} = J_{q}^{\,\prime,\,g}X_{q} +  J_{\xi_{1}}X_{1}^{g} + J_{\xi_{2}}X_{2}^{g}
\end{equation}
where
\begin{equation}\label{eq/TransportCoef/00}
\begin{array}{rl}
X_{q} &\equiv \displaystyle \frac{1}{T^{\ell}} - \frac{1}{T^{g}} \\\\
X_{j}^{g} & \equiv - \displaystyle \frac{1}{T^{\ell}}\left({\wmu}_{j}^{\ell}- {\wmu}_{j}^{g} + s_{j}^{g}(T^{\ell}-T^{g})\right), \quad j=1,2
\end{array}
\end{equation}
The resulting linear force-flux relations are
\begin{equation}\label{eq/TransportCoef/002a}
\begin{array}{rl}
X_{q} &= R_{qq}^{g}\,J_{q}^{\,\prime,\,g} + R_{q1}^{g}\,J_{\xi_{1}} + R_{q2}^{g}\,J_{\xi_{2}} \\\\
X_{1}^{g} &= R_{1q}^{g}\,J_{q}^{\,\prime,\,g} + R_{11}^{g}\,J_{\xi_{1}} + R_{12}^{g}\,J_{\xi_{2}}\\\\
X_{2}^{g} &= R_{2q}^{g}\,J_{q}^{\,\prime,\,g} + R_{21}^{g}\,J_{\xi_{i}} + R_{22}^{g}\,J_{\xi_{2}}
\end{array}
\end{equation}
or in the matrix notation
\begin{equation}\label{eq/TwoComponent/03}
\mathrm{X}^{g} \equiv \begin{pmatrix}[c] X_{q} \\ X_{1}^{g} \\ X_{2}^{g} \end{pmatrix}, \quad%
\mathrm{R}^{g} \equiv \begin{pmatrix} R_{qq}^{g} & R_{q1}^{g} & R_{q2}^{g} \\ R_{1q}^{g} & R_{11}^{g} & R_{12}^{g} \\ R_{2q}^{g} & R_{21}^{g} & R_{22}^{g} \end{pmatrix}, \quad%
\mathrm{J}^{g} \equiv \begin{pmatrix}[c] J_{q}^{\,\prime,\,g} \\ J_{\xi_1} \\ J_{\xi_2} \end{pmatrix} %
\end{equation}
we have
\begin{equation}\label{eq/TransportCoef/002}
\mathrm{X}^{g} = \mathrm{R}^{g}\spd\mathrm{J}^{g}
\end{equation}
The resistance matrix $\mathrm{R}^{g}$ satisfies the Onsager reciprocal relations, i.e. $R_{q1}^{g}=R_{1q}^{g}$, $R_{q2}^{g} = R_{2q}^{g}$, and $R_{21}^{g} =
R_{12}^{g}$.

In the above expressions for the entropy productions we used  the measurable heat flux \eqr{eq/ExcessEntropy/03} on the vapor side of the surface
$J_{q}^{\,\prime,\,g}$. One can similarly use the measurable heat flux on the liquid side of the surface $J_{q}^{\,\prime,\,\ell}$. The resulting resistance matrix
$\mathrm{R}^{\ell}$ differs from $\mathrm{R}^{g}$. We refer to \cite{kjelstrupbedeaux/heterogeneous} for the details of the alternative procedure.

\section{Kinetic theory}\label{sec/KineticTheory}

According to \cite[p. 180]{kjelstrupbedeaux/heterogeneous} kinetic theory gives the following expressions for the surface transport coefficients  for a two
component mixture
\begin{equation}\label{eq/KineticTheory/01}
\begin{array}{rl}
R_{qq}^{g} &= 4A\,\left\{1 + \displaystyle \frac{104}{25\pi}\left(\frac{w_{1}^{2}}{\varsigma_{1}} + \frac{w_{2}^{2}}{\varsigma_{2}}\right)\right\} \\\\
R_{qi}^{g} = R_{iq}^{g} &= 2RTA\,\left\{1 + \displaystyle \frac{16}{5\pi}\,\frac{w_{i}}{\varsigma_{i}}\right\}\,M_{i}^{-1} \\\\
R_{ij}^{g} &= (RT)^{2}\,A\,\left\{1 + 32\,\delta_{ij}\,\displaystyle\frac{1}{\varsigma_{i}}\left(\frac{1}{\sigma_{i}} + \frac{1}{\pi} -
\frac{3}{4}\right)\right\}\,M_{i}^{-1}\,M_{j}^{-1}
\end{array}
\end{equation}
where
\begin{equation}\label{eq/KineticTheory/01}
\begin{array}{rl}
A &\equiv \displaystyle 2^{-9/2}\,\sqrt{\pi}\,R\,(RT)^{-5/2} \big(c^{g}_{1}/\sqrt{M_{1}} + c^{g}_{2}/\sqrt{M_{2}} \big)^{-1}\\
\\
\varsigma_{i} &\equiv \displaystyle \big(c^{g}_{i}/\sqrt[4]{M_{i}}\big)/\big(c^{g}_{1}/\sqrt[4]{M_{1}} + c^{g}_{2}/\sqrt[4]{M_2}\big)\\
\\
w_{i} &\equiv \displaystyle \lambda_{i}/(\lambda_{1}+\lambda_{2})
\end{array}
\end{equation}
where $R$ is the universal gas constant, $\lambda_{i}$ and $c^{g}_{i}$ are the thermal conductivity and the gas coexistence concentration of the $i$-th component
respectively. $\sigma_{i}$ is the condensation coefficient of the $i$-th component, which are parameters in this theory, and $\delta_{ij}$ is the Kroneker symbol.
Furthermore, $M_{i}$, the molar mass of component $i$, appears in \eqr{eq/KineticTheory/01} to adopt the molar transfer coefficients used in
\cite{kjelstrupbedeaux/heterogeneous} to the mass transfer coefficients used in this paper. All these quantities and as a consequence the resistances are calculated
for a liquid and a vapor in coexistence at the temperature $T$ and the chemical potential difference $\mu_{12} \equiv \mu_1-\mu_2$ of the surface, see \cite[p.
180]{kjelstrupbedeaux/heterogeneous} in this context.

%
\section{Non-equilibrium continuous solution}\label{sec/NonEqSolution}

Assume we have the numerical solution for a particular non-equilibrium stationary state. That is we know all the fluxes $\mathrm{J}^{g}$ and forces $\mathrm{X}^{g}$
used in \eqr{eq/TransportCoef/002}: the constant fluxes are obtained directly from the non-equilibrium solution and the extrapolated bulk profiles are obtained
using the procedure described in \cite{glav/grad2}. We now consider the following problem: to determine the transport coefficients for the whole surface having the
non-equilibrium solution. This problem is, in a way, inverse to the common one, where one knows the resistances and, say, forces, and needs to determine the fluxes.
As one can see, \eqr{eq/TransportCoef/002a} has 9 unknown resistances\footnote{Solving the inverse problem we have to ensure the validity of the Onsager reciprocal
relations. This is one of the criteria to limit the size of the perturbation. This means that we have 9 independent resistances, but not 6.} and only 3 equations.
It is therefore not possible to determine all the transport coefficients uniquely having only one stationary state solution. In order to incorporate more equations
we need to consider other non-equilibrium stationary solutions which are independent of the previous. An important observation should be made here.

In \cite{glav/grad2} we have verified the validity of the hypothesis of local equilibrium of the surface. This implies, that the resistance matrix $\mathrm{R}^{g}$
is a function of thermodynamic parameters, say the temperature $T$ and the chemical potential difference $\mu_{12}$, of the surface: $\mathrm{R}^{g} =
\mathrm{R}^{g}(T^{s}, \mu_{12}^{s})$. In \cite{glav/grad2} we saw, that the temperature of the surface and the chemical potential difference of the surface depend
on both, the equilibrium temperature and the chemical potential difference, as well as on the size of the perturbation. Let $\ss$ indicate the size of a
non-equilibrium perturbation\footnote{Note, that a non-equilibrium state can be achieved by perturbing several independent quantities simultaneously. In this case
we have several perturbation parameters $\beta_{1},\ldots,\beta_{p}$. A measure $\ss$ is a norm of this $p$-dimensional vector of perturbations. The exact
expression for this norm is irrelevant, as soon as it goes to zero if and only if all $\beta_{1},\ldots,\beta_{p}$ go to zero.}, so that
\begin{equation}\label{eq/TransportCoef/11a}
\begin{array}{rl}
T^{s} &= T^{s}(T_{eq}, \mu_{12,\,eq}; \ss) \\
\mu_{12}^{s} &= \mu_{12}^{s}(T_{eq}, \mu_{12,\,eq}; \ss)
\end{array}
\end{equation}
Furthermore, $\mathrm{X}^{g} = \mathrm{X}^{g}(\ss)$ and $\mathrm{J}^{g} = \mathrm{J}^{g}(\ss)$. In order to be able to use several independent perturbations as a
source for the resistance coefficients, we must ensure that for all perturbations the temperature of the surface and the chemical potential of the surface are the
same. The simplest way to ensure this is to assume that $T^{s} \approx T_{eq}$ and $\mu_{12}^{s} \approx \mu_{12,\,eq}$. As is clear from
\eqr{eq/TransportCoef/11a}, this can be considered true if the perturbation rate $\ss$ is small enough. As we decrease $\ss$, the accuracy of this assumption
increases and in the limit $\ss \rightarrow 0$ it becomes exact.
It follows that
\begin{equation}\label{eq/TransportCoef/11b}
\mathrm{R}^{g} \equiv \mathrm{R}^{g}(T_{eq}, \mu_{12,\,eq}) = \lim_{\ss \rightarrow 0}\mathrm{R}^{g}(T_{eq}, \mu_{12,\,eq}; \ss)
\end{equation}
In practice there exists a particular size $\ss_{eq}$ of a perturbation, such that for all $\ss < \ss_{eq}$, $T^{s} \approx T_{eq}$ and $\mu_{12}^{s} \approx
\mu_{12,\,eq}$ with a satisfactory accuracy.

One should also note that the accuracy of a particular numerical procedure may impose a lower bound for the size of the the perturbation $\ss$ as well. All the
non-equilibrium profiles and therefore forces and fluxes are calculated by solving the system of differential equations numerically with some particular accuracy.
If a perturbation rate $\ss$ is lower then this accuracy, say $\ss_{num}$, then the data obtained from the numerical procedure are not reliable. We may therefore
use \eqr{eq/TransportCoef/002} only if the perturbation rate $\ss$ is in the range $\ss_{num} < \ss < \ss_{eq}$. The boundaries of this range should be established
empirically.


We determine the transport coefficients from two different methods: a "perturbation cell" method\footnote{This method was first used by Johannessen et. al. in
\cite{bedeaux/vdW/III} for one-component system. Here we discuss the grounds for the legitimacy of this procedure and generalize it to mixtures.} and an
experimental-like procedure. For ease of notation we will suppress the superscript $g$ in the rest of this section, as the procedure is the same for vapor and
liquid interfacial resistances.

\subsection{Perturbation cell}\label{sec/NonEqSolution/Cell}

Consider a stationary state which is perturbed from equilibrium by setting the temperature of the liquid\footnote{One should not confuse $T(\xl)$ with $T^{\ell}$.
The former is the actual temperature at $x=\xl$, i.e. at the box boundary on the liquid side. The latter is the temperature extrapolated from the liquid phase to
the interfacial region and calculated at $x=\xs$, i.e. at the dividing surface.} $T(\xl) = (1+\beta_{T})T_{eq}$, the pressure of the gas $p(\xg) =
(1+\beta_{p})p_{eq}$ and the mole fraction of the  liquid $\zeta^{\ell}(\xl) = (1+\beta_{\zeta})\zeta^{\ell}_{eq}$ independently. The resulting non-equilibrium
state is therefore a function of the parameters $\beta$:
\begin{equation}\label{eq/TwoComponent/04}
\mathrm{X}(\beta_{T}, \beta_{p}, \beta_{\zeta}) = \mathrm{R}(T_{eq}, \mu_{12,\,eq})\spd\mathrm{J}(\beta_{T}, \beta_{p}, \beta_{\zeta})
\end{equation}
where $\mathrm{X}$, $\mathrm{J}$ and $\mathrm{R}$ are given by \eqr{eq/TwoComponent/03}. Consider the following set of 8 independent non-equilibrium perturbations:
\begin{equation}\label{eq/TwoComponent/07}
\begin{array}{rrrrcrcrrrr}
\mathrm{X}(&\beta,& \beta,& \beta) &=& \mathrm{R}(T_{eq}, \mu_{12,\,eq})&\spd&\mathrm{J}(&\beta,& \beta,& \beta) \\
\mathrm{X}(&\beta,& -\beta,& \beta) &=& \mathrm{R}(T_{eq}, \mu_{12,\,eq})&\spd&\mathrm{J}(&\beta,& -\beta,& \beta) \\
\mathrm{X}(&-\beta,& \beta,& \beta) &=& \mathrm{R}(T_{eq}, \mu_{12,\,eq})&\spd&\mathrm{J}(&-\beta,& \beta,& \beta) \\
\mathrm{X}(&-\beta,& -\beta,& \beta) &=& \mathrm{R}(T_{eq}, \mu_{12,\,eq})&\spd&\mathrm{J}(&-\beta,& -\beta,& \beta) \\
\mathrm{X}(&\beta,& \beta,& -\beta) &=& \mathrm{R}(T_{eq}, \mu_{12,\,eq})&\spd&\mathrm{J}(&\beta,& \beta,& -\beta) \\
\mathrm{X}(&\beta,& -\beta,& -\beta) &=& \mathrm{R}(T_{eq}, \mu_{12,\,eq})&\spd&\mathrm{J}(&\beta,& -\beta,& -\beta) \\
\mathrm{X}(&-\beta,& \beta,& -\beta) &=& \mathrm{R}(T_{eq}, \mu_{12,\,eq})&\spd&\mathrm{J}(&-\beta,& \beta,& -\beta) \\
\mathrm{X}(&-\beta,& -\beta,& -\beta) &=& \mathrm{R}(T_{eq}, \mu_{12,\,eq})&\spd&\mathrm{J}(&-\beta,& -\beta,& -\beta) \\
\end{array}
\end{equation}
Consider now the $3 \times 8$ matrices $\mathfrak{X}$ and $\mathfrak{J}$ which contain 8 column vectors $\mathrm{X}$ and $\mathrm{J}$ respectively for each
non-equilibrium perturbation specified above. For these perturbations $\mathfrak{X} = \mathfrak{X}(\beta)$ and $\mathfrak{J} = \mathfrak{J}(\beta)$ are functions
only of one parameter $\beta$. It follows from \eqr{eq/TwoComponent/07} that
\begin{equation}\label{eq/TwoComponent/08}
\mathfrak{X}(\beta) = \mathrm{R}(T_{eq}, \mu_{12,\,eq}) \spd \mathfrak{J}(\beta)
\end{equation}
where $\beta$ should be in the appropriate range, as discussed above. In \appr{sec/Perturbation} we discuss the method to obtain this range. From
\eqr{eq/TwoComponent/08} we obtain
\begin{equation}\label{eq/TwoComponent/09}
\mathrm{R}(T_{eq}, \mu_{12,\,eq}) = \left(\mathfrak{X}(\beta)\spd\mathfrak{J}^{T}(\beta)\right)\spd \left(\mathfrak{J}(\beta)\spd\mathfrak{J}^{T}(\beta)\right)^{-1} \\
\end{equation}
where superscript ${}^{T}$ means the matrix transpose and ${}^{-1}$ means the inverted matrix.

We note, that in order to obtain the resistance matrix $\mathrm{R}$ uniquely, it is sufficient in principle to impose any 3 non-equilibrium perturbations which have
sufficiently small perturbation parameters $\beta_{T}$, $\beta_{p}$ and $\beta_{\zeta}$. This would give us $3 \times 3 = 9$ independent equations. The method
presented above makes the resistance matrix converge to $\mathrm{R}(T_{eq},\mu_{12,\,eq})$ as fast as $\beta^{2}$ goes to zero, however. This is achieved by using 8
symmetric perturbations at the "corners" of a three-dimensional "perturbation cell", so changing $\beta $ to $-\beta $ does not change the "perturbation cell" and
the resulting $\mathrm{R}$.

Because of using 8 perturbations instead of 3, there are 5 superfluous perturbations which make the system of equations \eqref{eq/TwoComponent/08} to be
overdetermined. Contracting both sides of \eqr{eq/TwoComponent/08} with $\mathfrak{J}^{T}$ we actually average all the perturbations which are spread around
$T_{eq}$ and $\mu_{12,\,eq}$ in the least square sense. As the components of $\mathfrak{J}$ matrix are linearly independent, this guaranteers the matrix
$\mathfrak{J}\spd\mathfrak{J}^{T}$ to be invertible. Thus, the inverse matrix $(\mathfrak{J}\spd\mathfrak{J}^{T})^{-1}$ exists and \eqr{eq/TwoComponent/09} is
mathematically legitimate. In the numerical procedure the expression on the right hand side of \eqr{eq/TwoComponent/09} is obtained using Matlab matrix division
$/$.

\subsection{Experiment-like procedure}\label{sec/NonEqSolution/Experiment}

In experiments it is convenient to measure the corresponding coefficients by keeping zero mass fluxes through the system. It is also convenient  to work with the
total mass flux $J_{m} = J_{\xi_{1}} + J_{\xi_{2}}$ and the flux of one of the components $J_{\xi} \equiv J_{\xi_{1}} $, rather then with fluxes of each component
separately\footnote{One of the reasons for this is that it is hard to make only $J_{\xi_{1}} = 0$, keeping $J_{\xi_{2}}$ finite.}, $J_{\xi_{1}}$ and $J_{\xi_{2}}$.
The excess entropy production \eqr{eq/TwoComponent/00} can be therefore written as
\begin{equation}\label{eq/TwoComponent/10}
{\sigma}_{s} = J_{q}^{\,\prime}X_{q} +  J_{\xi}X_{\xi} + J_{m}X_{m}
\end{equation}
where $X_{\xi} \equiv X_{1}-X_{2}$ and $X_{m} \equiv X_{2}$. The resulting force-flux relations \eqref{eq/TransportCoef/002} have the following terms
\begin{equation}\label{eq/TwoComponent/03a}
\mathrm{X} \equiv \begin{pmatrix}[c] X_{q} \\ X_{\xi} \\ X_{m} \end{pmatrix}, \quad%
\mathrm{R} \equiv \begin{pmatrix} R_{qq} & R_{q\xi} & R_{qm} \\ R_{\xi q} & R_{\xi\xi} & R_{\xi m} \\ R_{mq} & R_{m\xi} & R_{mm} \end{pmatrix}, \quad%
\mathrm{J} \equiv \begin{pmatrix}[c] J_{q}^{\,\prime} \\ J_{\xi} \\ J_{m} \end{pmatrix} %
\end{equation}
where the resistances for different force definitions are related as
\begin{equation}\label{eq/TwoComponent/12}
\begin{pmatrix}
R_{qq} & R_{q1} & R_{q2} \\
R_{1q} & R_{11} & R_{12} \\
R_{2q} & R_{21} & R_{22} \end{pmatrix} =%
\begin{pmatrix}
R_{qq} \;&\; R_{q\xi} \!+\! R_{qm} \;&\; R_{qm} \\
R_{\xi q} \!+\! R_{mq} \;&\; R_{mm} \!+\! R_{\xi\xi}\!-\!R_{m\xi} \!-\! R_{\xi m} \;&\; R_{mm} \!-\! R_{\xi m} \\
R_{mq} \;&\; R_{mm} \!-\! R_{m\xi} \;&\; R_{mm}
\end{pmatrix}
\end{equation}
%

Consider a stationary state which is perturbed from equilibrium by setting the temperature of the liquid $T(\xl) = (1+\beta)T_{eq}$. The perturbation parameter
$\beta$ is a small number. The second perturbation constraint we impose is either $J_{\xi} = 0$ or $\zeta^{\ell}(\xl) = \zeta^{\ell}_{eq}$ and we introduce the
perturbation parameter $\nu_{\xi}$ which is 0 in the former case and 1 in the latter one, which will be used as a subscript. The third perturbation condition is
either $J_{m} = 0$ or $p(\xg) = p_{eq}$ and the corresponding perturbation parameter $\nu_{m}$ is 0 or 1 respectively, which will be used as a subscript. The
resulting non-equilibrium state is therefore a function of 3 parameters:
\begin{equation}\label{eq/TwoComponent/13}
\mathrm{X}_{\nu_{\xi}, \nu_{m}}(\beta) = \mathrm{R}(T_{eq}, \mu_{12,\,eq})\spd\mathrm{J}_{\nu_{\xi}, \nu_{m}}(\beta)
\end{equation}
where $\mathrm{X}$, $\mathrm{J}$ and $\mathrm{R}$ are given by \eqr{eq/TwoComponent/03a}.

Consider the following set of 3 independent non-equilibrium perturbations:
\begin{equation}\label{eq/TwoComponent/15}
\begin{array}{rl}
\bigVarea \mathrm{X}_{00}(\beta) &= \mathrm{R}(T_{eq}, \mu_{12,\,eq})\spd\mathrm{J}_{00}(\beta)\\
\bigVarea \mathrm{X}_{10}(\beta) &= \mathrm{R}(T_{eq}, \mu_{12,\,eq})\spd\mathrm{J}_{10}(\beta)\\
\bigVarea \mathrm{X}_{11}(\beta) &= \mathrm{R}(T_{eq}, \mu_{12,\,eq})\spd\mathrm{J}_{11}(\beta)\\
\end{array}
\end{equation}
Further on for simplicity we will suppress arguments $\beta$ and $(T_{eq}, \mu_{12,\,eq})$.

%
\begin{subequations}\label{eq/TwoComponent/16}
From the first of \eqr{eq/TwoComponent/15} we find
\begin{equation}\label{eq/TwoComponent/16a}
\begin{array}{rcrl}
\BigVarea R_{qq}    &=& X_{q,\;\,00}    &/\, J_{q,\,00}^{\,\prime} \\
\BigVarea R_{\xi q} &=& X_{\xi,\;\,00}  &/\, J_{q,\,00}^{\,\prime}\\
\BigVarea R_{mq}    &=& X_{m,\,00}      &/\, J_{q,\,00}^{\,\prime}\\
\end{array}
\end{equation}
From the second of \eqr{eq/TwoComponent/15} we find
\begin{equation}\label{eq/TransportCoef/16b}
\begin{array}{rcrl}
\BigVarea R_{q\xi}   &=& \left(X_{q,\;\,10}     - R_{qq}   \,J_{q,\,10}^{\,\prime}\right)    &/\, J_{\xi,\,10} \\
\BigVarea R_{\xi\xi} &=& \left(X_{\xi,\;\,10}   - R_{\xi q}\,J_{q,\,10}^{\,\prime}\right)  &/\, J_{\xi,\,10} \\
\BigVarea R_{m\xi}   &=& \left(X_{m,\,10}       - R_{mq}   \,J_{q,\,10}^{\,\prime}\right)    &/\, J_{\xi,\,10} \\
\end{array}
\end{equation}
The values $\mathrm{X}_{10}$ and $\mathrm{J}_{10}$ are found directly from the calculations and the values of $R_{qq}$, $R_{\xi q}$ and $R_{mq}$ are those which are
found in \eqr{eq/TwoComponent/16a}, given that the perturbation rate $\beta$ is small enough. From the third of \eqr{eq/TwoComponent/15} we find
\begin{equation}\label{eq/TransportCoef/16c}
\begin{array}{rcrl}
\BigVarea R_{qm}    &=& \left(X_{q,\;\,11}\;   - R_{qq}     \,J_{q,\,11}^{\,\prime} - R_{q\xi}  \,J_{\xi,\,11} \right) &/\, J_{m,\,11} \\
\BigVarea R_{\xi m} &=& \left(X_{\xi,\;\,11}\; - R_{\xi q}  \,J_{q,\,11}^{\,\prime} - R_{\xi\xi}\,J_{\xi,\,11} \right) &/\, J_{m,\,11} \\
\BigVarea R_{mm}    &=& \left(X_{m,\,11}       - R_{mq}     \,J_{q,\,11}^{\,\prime} - R_{m\xi}  \,J_{\xi,\,11} \right) &/\, J_{m,\,11} \\
\end{array}
\end{equation}
Again, all the quantities on the right hand side of \eqr{eq/TransportCoef/16c} are known and we therefore can find the remaining resistivities.
\end{subequations}
\section{Integral relations}\label{sec/IntegralRelations}

In \cite{glav/tc1} we have established the general approach to derive integral relations between the surface resistances and local resistivity profiles. In this
section we apply it to find the relations between the resistances $\mathrm{R}$ used in \eqr{eq/TransportCoef/002} and \eqr{eq/TwoComponent/03} and local
resistivities $\mathrm{r}$. Using the method described in \cite{glav/tc1} we find
\begin{equation}\label{eq/Binary/04}
\begin{array}{rl}
R^{\,\prime\,g}_{qq} =& \mathfrak{E}\,\{r_{qq}\} \\\\
R^{\,\prime\,g}_{q1} =& \mathfrak{E}\,\{r_{qq}(h-h_{1}^{g}) + r_{q1}\,\xi_{2} \} \\\\
R^{\,\prime\,g}_{q2} =& \mathfrak{E}\,\{r_{qq}(h-h_{2}^{g}) - r_{q1}\,\xi_{1} \} \\\\
R^{\,\prime\,g}_{11} =& \mathfrak{E}\,\{r_{qq}(h-h_{1}^{g})^{2} + 2r_{q1}\,\xi_{2}\,(h-h_{1}^{g}) + r_{11}\,\xi_{2}^{2} \} \\\\
R^{\,\prime\,g}_{12} =& \mathfrak{E}\,\{r_{qq}(h-h_{1}^{g})(h-h_{2}^{g}) + r_{q1}(\xi_{2}\,(h-h_{2}^{g}) - \xi_{1}\,(h-h_{1}^{g})) - r_{11}\,\xi_{1}\,\xi_{2} \} \\\\
R^{\,\prime\,g}_{22} =& \mathfrak{E}\,\{r_{qq}(h-h_{2}^{g})^{2} - 2r_{q1}\,\xi_{1}\,(h-h_{2}^{g}) + r_{11}\,\xi_{1}^{2} \} \\\\
\end{array}
\end{equation}
where the operator $\mathfrak{E}$ is defined as
\begin{equation}\label{eq/Resistivities/04}
\mathfrak{E}\{\phi\}(\xs) \equiv \int_{\displaystyle \xgsb}^{\displaystyle \xlsb}{dx\left[\phi(x) - \phi^{g}(x)\,\Theta(\xs-x) -
\phi^{\ell}(s)\,\Theta(x-\xs)\right]}
\end{equation}%
where $\phi^{g}$ and $\phi^{\ell}$ are extrapolated from the gas and liquid respectively profiles of $\phi$, while $\xgsb$ and $\xlsb$ are the surface boundaries.

This method requires the equilibrium profiles for the enthalpy $h(x)$ and the mass fraction $\xi(x)$ across the interface. Both of them could be easily obtained
from the equilibrium square gradient model, see \cite{glav/grad2} for details. In contrast to the methods in \secr{sec/NonEqSolution}, this requires calculating
only the equilibrium profiles, but not the non-equilibrium ones, which is a much easier calculation.

The integral relations also require the local resistivity profiles $r_{qq}(x)$, $r_{q1}(x)$, and $r_{11}(x)$ across the interface, which were modeled in the square
gradient theory as
%
\begin{equation}\label{eq/Equations/Phenomenological/03}
\begin{array}{rl}
r_{qq}(x) &= r_{qq}^{g} + (r_{qq}^{\ell}-r_{qq}^{g})\,q_{0}(x) + \alpha_{qq}(r_{qq}^{\ell}+r_{qq}^{g})\,q_{1}(x)\\
\\
r_{q1}(x) &= r_{q1}^{g} + (r_{q1}^{\ell}-r_{q1}^{g})\,q_{0}(x) + \alpha_{q1}(r_{q1}^{\ell}+r_{q1}^{g})\,q_{1}(x)\\
\\
r_{11}(x) &= r_{11}^{g} + (r_{11}^{\ell}-r_{11}^{g})\,q_{0}(x) + \alpha_{11}(r_{11}^{\ell}+r_{11}^{g})\,q_{1}(x)\\
\end{array}
\end{equation}
where $q_{0}(x)$ and $q_{1}(x)$ are modulatory curves for resistivity profiles which depend only on density profiles and their first derivatives. We refer for the
details to \cite{glav/grad2}. $q_{0}(x)$ is a smooth $\arctan$-like function which changes its value from 0 to 1 within the range $[\xgsb; \xlsb]$ and $q_{1}(x)$ is
zero on the boundaries of the $[\xgsb; \xlsb]$ interval and has a peak proportional to the square gradient of the density inside this interval. Thus, the first two
terms in each expression for the resistivity represents a smooth transitions from the gas bulk resistivity to the liquid bulk resistivity, while the third term
represents a peak in the resistivity proportional to the square gradient of the density. For each resistivity profile $r^{g}$ and $r^{\ell}$ are the equilibrium
coexistence resistivities of the gas and liquid phase respectively. They are related to the measurable transport coefficients such as heat conductivity, the
diffusion coefficient and the Soret coefficient.

The square gradient model used 3 adjustable parameters $\alpha_{qq}$, $\alpha_{q1}$, $\alpha_{11}$ which control the size of the peak in the resistivity profiles in
the interfacial region. The interfacial resistance coefficients $\mathrm{R}$ will therefore depend on these coefficients, $\mathrm{R} = \mathrm{R}(\alpha_{qq},
\alpha_{q1}, \alpha_{11})$, which we will investigate.

\section{Results}\label{sec/Results}

We consider a binary mixture of cyclohexane and $n$-hexane, as we did in \cite{glav/grad2}.

We find in \appr{sec/Perturbation} that $\beta = 2\spd10^{-4}$ is an optimum perturbation rate both in the "perturbation cell" and "experimental-like" methods, see
\secr{sec/NonEqSolution}. We have verified that both these methods lead to essentially the same values of the resistance coefficients. The numbers given below are
taken from the "perturbation cell" method.

Furthermore in \appr{sec/Consistency}, we find the range of adjustable amplitudes $\alpha_{qq}$, $\alpha_{1q}$ and $\alpha_{11}$, for which the description is
thermodynamically consistent. We find that $\alpha_{qq} \sim 10$, $\alpha_{11} \sim 1$. The value of $\alpha_{1q}$ is found to be irrelevant.

In this section we suppress the superscript $g$ for the resistances for ease of notation.


%
\subsection{Comparison to kinetic theory}

In this subsection we investigate the values of parameters  $\alpha_{qq}$, $\alpha_{1q}$, $\alpha_{11}$ which makes the coefficients agree  with the kinetic theory
coefficients. We do it for $\beta$ = 2e-4 as this perturbation rate gives the most accurate results. Furthermore we use the temperature $T_{eq} = 330$ K and
chemical potential difference $\mu_{12,\,eq} = 700$ J/mol. The values of parameters, used for kinetic theory are the same, as we use in our calculations.
Particularly, the heat conductivities are $\lambda_{1} = 0.0140$ W/(m K) and $\lambda_{2} = 0.0157$ W/(m K), $M_{1} = 84.162$  g/mol and $M_{2} = 86.178$ g/mol.

We found that a variation of $\alpha_{1q}$ from 0 to 10 makes the diagonal coefficients vary about 1 \% and the cross coefficients vary not more then 5 \%. As the
variation of $\alpha_{1q}$ is quite substantial, the variation in the coefficients which it induces is negligible. We therefore take $\alpha_{1q} = 0$ in all
further analysis.

Let us use subscript $pc$ for the resistivity matrix obtained from the "perturbation cell" method and subscript $kin$ for the resistivity matrix obtained from
kinetic theory. For the above parameters $R_{qq, pc} = 2.96792 \times 10^{-11}$. We found that $R_{qq, kin}$ is practically independent on $\alpha_{11}$ while it
depends linearly on $\alpha_{qq}$, see \figr{R_{qq}-A_{qq}}. One can see from the plot, that they are the same for $\alpha_{qq} \approx 9$.
\begin{figure}[hbt!]
\centering
\includegraphics[scale=\profilescale]{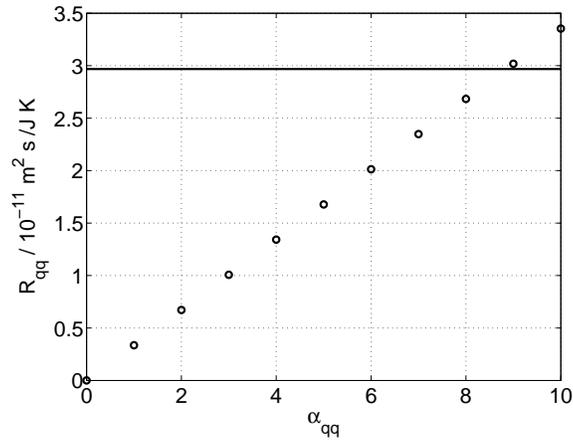}
\caption{Dependence of $R_{qq}$ on $\alpha_{qq}$ obtained by "perturbation cell" method at $T_{eq} = 330$ and $\mu_{12,\,eq} = 700$ for $\alpha_{1q} = 0$ and
$\alpha_{11} = 1$. $R_{qq,\,kin}$ is drawn as a constant line.}\label{R_{qq}-A_{qq}}
\end{figure}

The diagonal coefficients $R_{11, pc}$ and $R_{22, pc}$ depend both on $\alpha_{qq}$ and $\alpha_{11}$. Since we have found the value of $\alpha_{qq}$ already, we
will further investigate the dependence of $R_{11, pc}$ and $R_{22, pc}$ using this value of $\alpha_{qq}$ and varying only $\alpha_{11}$. The diagonal coefficients
$R_{11, kin}$ and $R_{22, kin}$ depend, in their turn, on the condensation coefficients $\sigma_{1}$ and $\sigma_{2}$ respectively. We plot this dependence in the
same plot with the dependency of $R_{ii, pc}$ ($i=1;2$) on $\alpha_{11}$, see \figr{R_{ii}-A_{11}}. The dependence of $R_{ii, pc}$ on $\alpha_{11}$ is given by the
dotted line with the values of $\alpha_{11}$ drawn on the bottom $x$-axes. The dependence of $R_{ii, kin}$ on $\sigma_{i}$ is given by the solid line with the
values $\sigma_{i}$ drawn on the top $x$-axes.
\begin{figure}
\centering
\subfigure
{\includegraphics[scale=\profilescale]{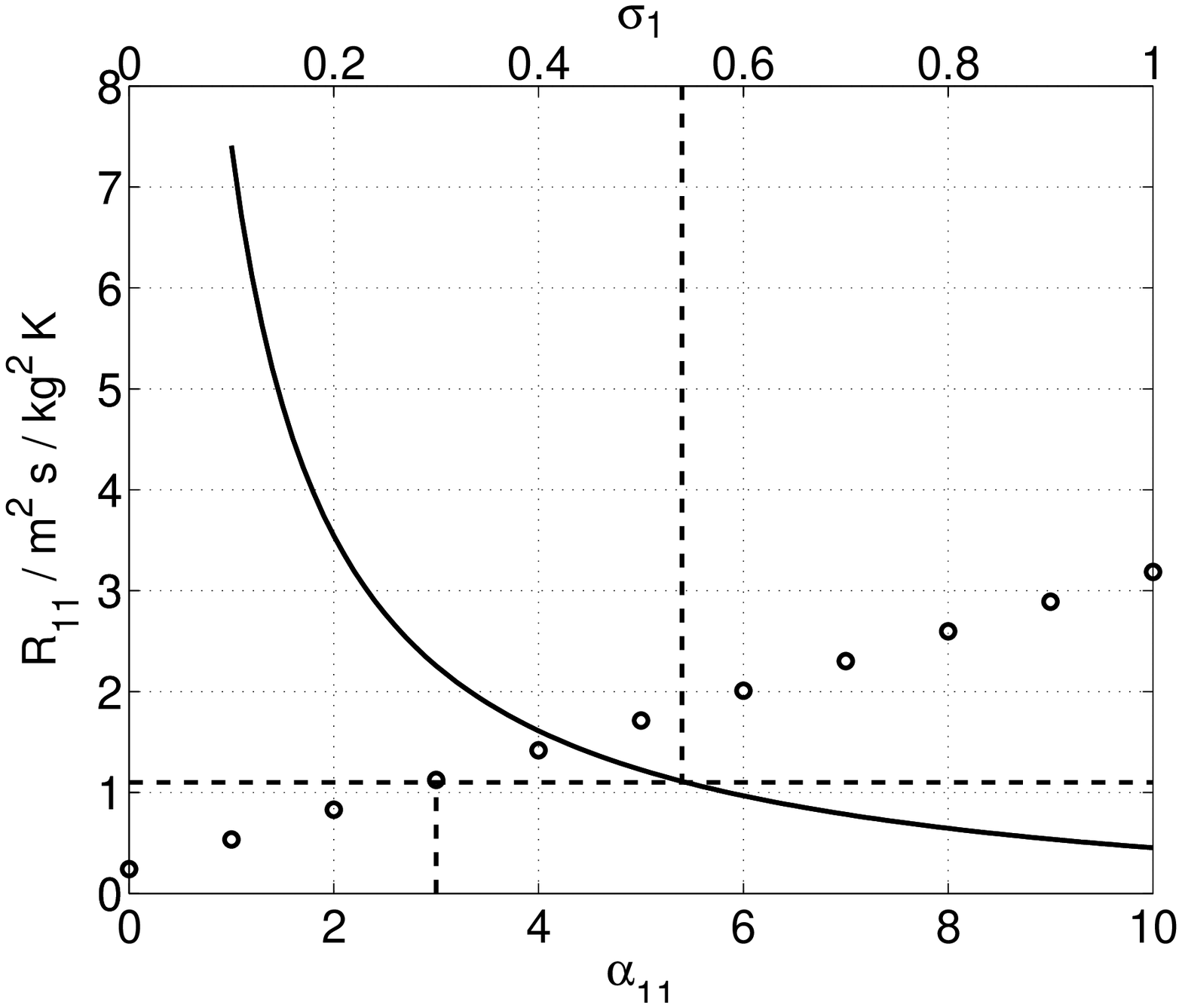}\label{R_{11}-A_{11},S_{1}} } %
\subfigure
{\includegraphics[scale=\profilescale]{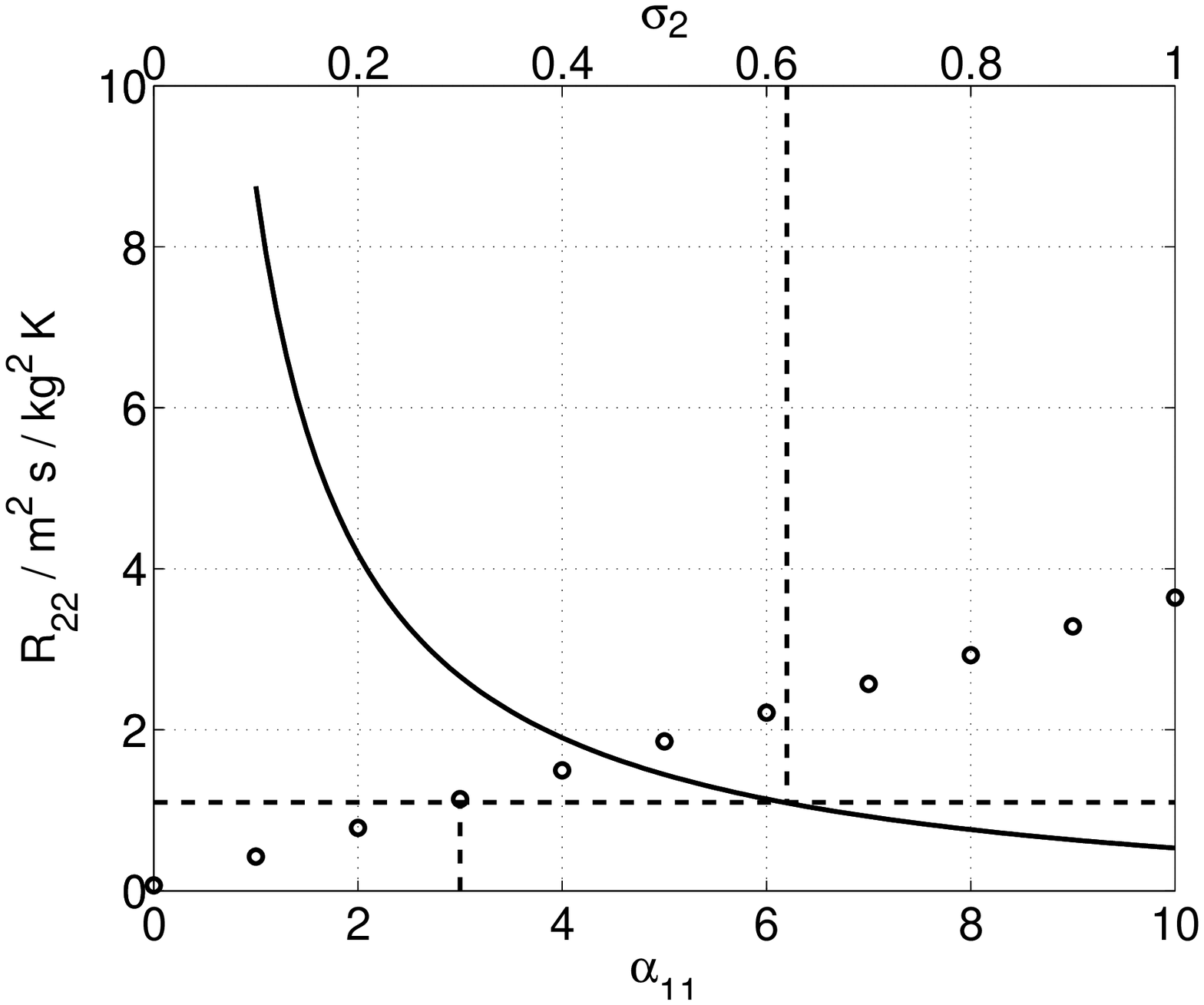} \label{R_{22}-A_{11,S_{2}}} } %
\caption{Dependence of $R_{11,\,pc}$ and $R_{22,\,pc}$ on $\alpha_{11}$ (dots, bottom axes) and $R_{11,\,kin}$ and $R_{22,\,kin}$ on $\sigma_{1}$ and $\sigma_{2}$
(curve, top axes), respectively. Data are obtained at $T_{eq} = 330$ and $\mu_{12,\,eq} = 700$ for $\alpha_{qq} = 9$ and $\alpha_{1q} = 0$.}\label{R_{ii}-A_{11}}
\end{figure}

Consider a particular value $R_{ii,\,0}$ of the diagonal coefficient $R_{ii}$, where $i$ is either 1 or 2, which is indicated by a horizontal  dashed line on a
figure. To find the value of $\alpha_{11}$ for which $R_{ii, pc} = R_{ii,\,0}$ we draw a perpendicular from the point where it crosses the dotted line to the bottom
axes. To find the value of $\sigma_{i}$ for which $R_{ii, kin} = R_{ii,\,0}$ we draw a perpendicular from the point where the horizontal dashed line crosses the
solid line to the top axes. For instance, the value $R_{22,\,0} = 1.1$ corresponds to $\alpha_{11} = 3$ and $\sigma_{2} = 0.62$. The value $\alpha_{11} = 3$, in its
turn, gives $R_{11,\,0} = 1.1$ which corresponds to $\sigma_{1} = 0.54$.

One may start by specifying $\alpha_{11}$, rather then $R_{ii,\,0}$, to find $\sigma_{1}$ and $\sigma_{2}$. Then we draw a perpendicular from  the bottom axes until
it crosses the dotted line, which gives the value of $R_{ii, pc}$. Given the value of $R_{ii, kin}$ to be the same, we find the value of $\sigma_{i}$ as described
above. For the above example $\alpha_{11} = 3$ corresponds to $\sigma_{1} = 0.54$ and $\sigma_{2} = 0.62$. We see, that we may not specify both $\sigma_{1}$ and
$\sigma_{2}$ independently: they must have the values which both correspond to the same $\alpha_{11}$. For similar components, like those we are interested in,
$\sigma_{1}$ and $\sigma_{2}$ should not differ much from each other, and therefore $\alpha_{11}$, a coefficient which is related to the diffusion of one component
through the other, should reflect this difference.


Having the diagonal coefficient mapped we have the parameters $\alpha_{qq}$ and $\alpha_{11}$ defined uniquely (and taking into account  that $\alpha_{1q}$ has
negligible effect), as well as $\sigma_{1}$ and $\sigma_{2}$ for kinetic theory. We now compare the values of the cross coefficients given by "perturbation cell"
method and kinetic theory.
\begin{longtable}{l@{\qquad}l@{\qquad}l@{\qquad}l@{\qquad}l@{\qquad}l@{\qquad}l} %
\caption{Gas-side transport coefficients obtained from kinetic theory and by "perturbation cell" method at $T_{eq} = 330$ and $\mu_{12,\,eq} = 700$ for $\beta = 0.0002$.}\label{tbl/Coeff/Kn-Pc-4-9-0-3}\\%
\hline %
parameters   & $R_{qq}$  & $R_{11}$  & $R_{22}$  & $R_{q1}$  & $R_{q2}$  & $R_{12}$  \\%
\hline %
\begin{tabular}{l} $\sigma_{1}=0.54$ \\ $\sigma_{2}=0.62$ \end{tabular} & 2.96792e-011   & 1.11091   & 1.09136   & 3.82826e-007     & 4.41483e-007     & 0.0130511      \\%
\begin{tabular}{l} $\alpha_{qq}=9$ \\ $\alpha_{1q}=0$ \\ $\alpha_{11}=3$\end{tabular} & 3.01874e-011     & 1.12461   & 1.13991   & 2.31477e-006     & 2.27003e-006     & -0.816559      \\%
\hline %
\end{longtable} %
One can see from \tblr{tbl/Coeff/Kn-Pc-4-9-0-3} that while the diagonal coefficients are the same\footnote{One should not expect exact compatibility between kinetic
theory, which is most appropriate for gases with short range potentials, and the gradient theory, which is most appropriate for fluids with long range potentials.
The purpose of this comparison in not to determine the exact values of adjustable parameters, but to show that it is possible to match coefficients in the two
theories and to show the typical values of the parameters.}, the cross coefficients we find are an order of magnitude larger than those found by kinetic theory.
$R_{12}$ even has a different sign.

\subsection{Temperature and chemical potential difference dependence}

\begin{figure}[hbt!]
\centering
\includegraphics[scale=\profilescale]{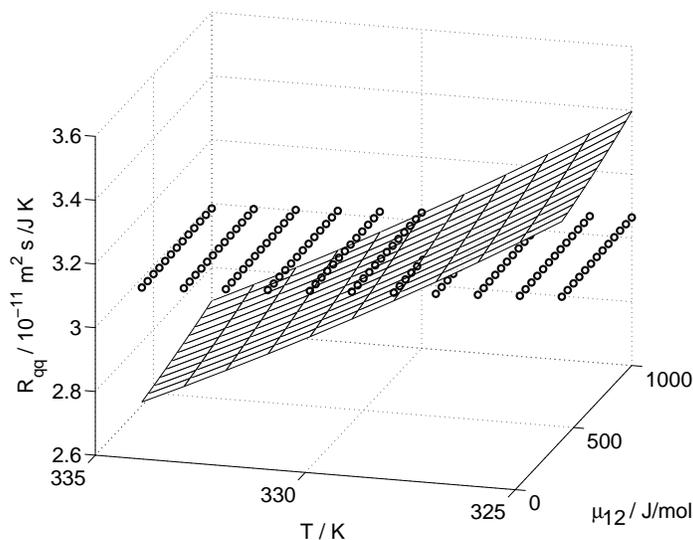}
\caption{Dependence of $R_{qq}$ on $T$ and $\mu_{12}$ obtained from kinetic theory for $\sigma_{1}=0.54$ and $\sigma_{2}=0.62$ (plane) and by "perturbation cell"
method for $\alpha_{qq} = 9$, $\alpha_{1q} = 0$ and $\alpha_{11} = 3$ (points).}\label{Rqq}
\end{figure}

In this subsection we investigate the dependence of the resistivity coefficients on the temperature and the chemical potential difference.  On \figsr{Rqq}{R11} we
plot the these dependencies for $R_{qq}$, $R_{q1}$ and $R_{11}$ coefficients obtained from kinetic theory and "perturbation cell" method for the range of
temperatures $[325,\ldots,335]$ and for the range of chemical potential differences $[400,\ldots,1000]$.
\begin{figure}[hbt!]
\centering
\includegraphics[scale=\profilescale]{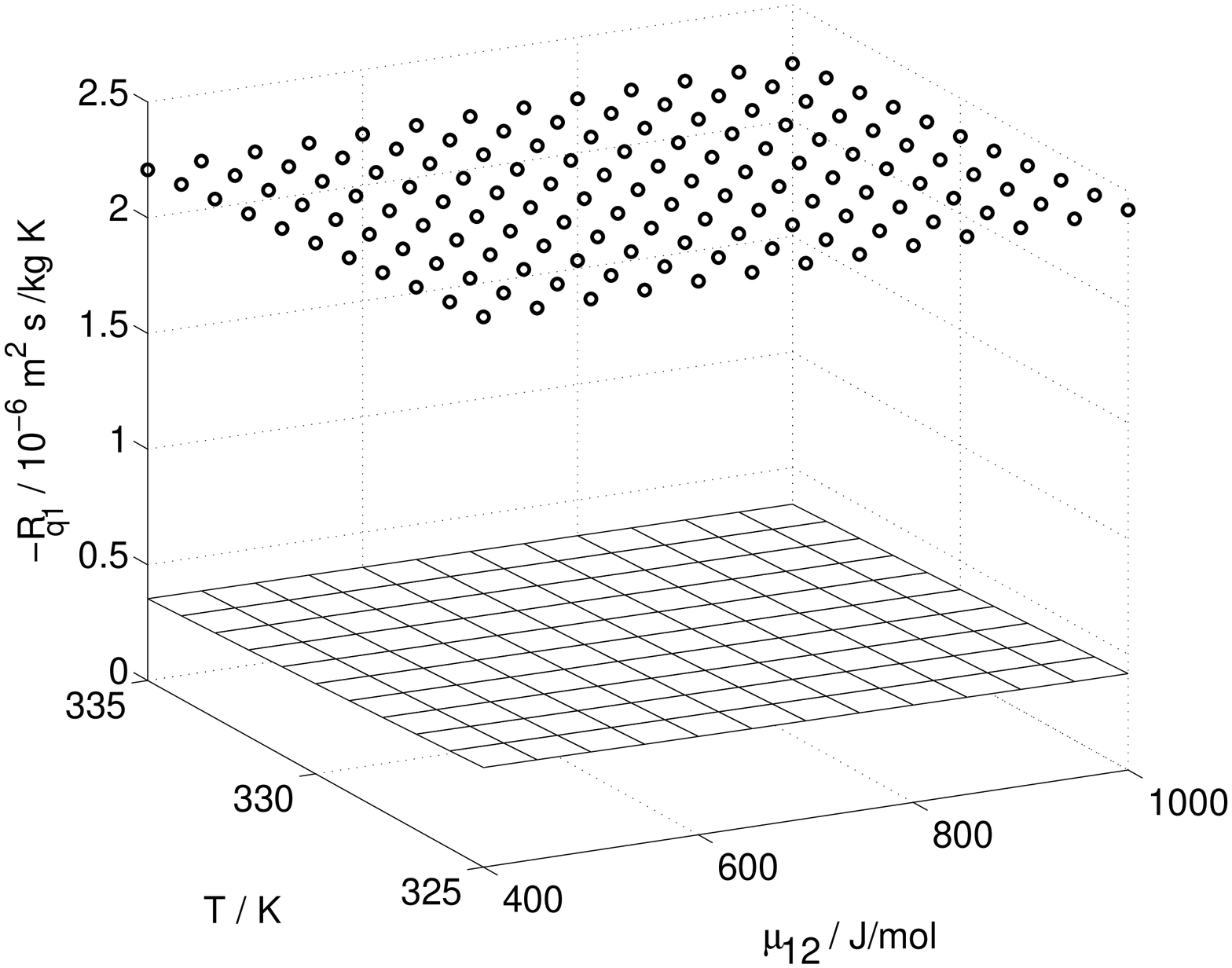}
\caption{Dependence of $R_{q1}$ on $T$ and $\mu_{12}$ obtained from kinetic theory for $\sigma_{1}=0.54$ and $\sigma_{2}=0.62$ (plane) and by "perturbation cell"
method for $\alpha_{qq} = 9$, $\alpha_{1q} = 0$ and $\alpha_{11} = 3$ (points).}\label{Rq1}
\end{figure}
\begin{figure}[hbt!]
\centering
\includegraphics[scale=\profilescale]{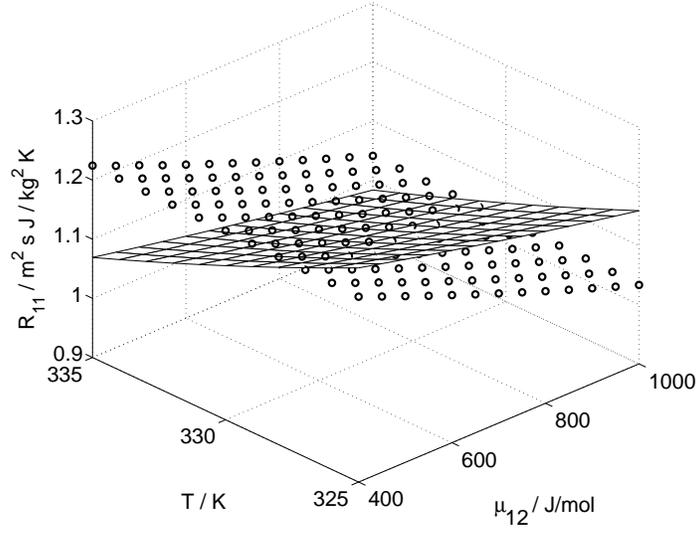}
\caption{Dependence of $R_{11}$ on $T$ and $\mu_{12}$ obtained from kinetic theory for $\sigma_{1}=0.54$ and $\sigma_{2}=0.62$ (plane) and by "perturbation cell"
method for $\alpha_{qq} = 9$, $\alpha_{1q} = 0$ and $\alpha_{11} = 3$ (points).}\label{R11}
\end{figure}

The domain of $T$ and $\mu_{12}$ is not big, so the dependence on them is linear, as expected.

\subsection{Validity of integral relations}

We compare the resistances found from numerical procedure to the values obtained from \eqr{eq/Binary/04}. The relative difference between them is almost the same
within the range of temperatures and chemical potential differences considered: $T=\{325,\cdots,335\}$ and $\mu_{12}=\{400,\cdots,1000\}$. In
\tblr{tbl/Diff/Pc-Ir-4-9-0-3} we give the relative errors for the resistances both for the case that we use the measurable heat fluxes on the vapor and on the
liquid side. We refer to \cite{kjelstrupbedeaux/heterogeneous} for details of the definition of the resistances using the measurable heat flux on the liquid side.

\begin{longtable}{l@{\qquad}l@{\qquad}l@{\qquad}l@{\qquad}l@{\qquad}l@{\qquad}l@{\qquad}l@{\qquad}l@{\qquad}l}%
\caption{Relative error in percent between the gas- and liquid- side coefficients obtained by "perturbation cell" and "integral relations" methods at $T_{eq} = 330$ and $\mu_{12,\,eq} = 700$ for $\beta = 0.0002$ and $\alpha_{qq} = 9$, $\alpha_{1q} = 0$, $\alpha_{11} = 3$ .} \label{tbl/Diff/Pc-Ir-4-9-0-3}\\%
\hline %
phase & $R_{qq}$     & $R_{11}$  & $R_{22}$  & $R_{q1}$  & $R_{q2}$  & $R_{12}$ \\%
\hline %
gas     & 0.019090   & 0.064642  & 0.058851  & 0.020649  & 0.020680  & 0.097096  \\%
liquid  & 0.019090   & 0.006266  & 0.000432  & 0.036270  & 0.034886  & 6.233983  \\%
\hline %
\end{longtable} %
The relative differences are not more then a few promille. It is larger only for $R_{12}^{\ell}$ which is discussed below.

\section{Discussion and conclusions}\label{sec/Discussion}

In this paper we have studied stationary transport of heat and mass through the liquid-vapor interface in a mixture. We used the expression for the excess entropy
production of a surface derived from the continuous description \cite{glav/tc1}, which is identical to the one derived directly for the discrete description using
the property of local equilibrium \cite{kjelstrupbedeaux/heterogeneous}. This makes it possible to give the linear force-flux relations for this case. These
relations involve the interfacial resistances, which were the main focus of interest in this paper. Given the numerical solutions of the non-equilibrium square
gradient model we were able to calculate these coefficients directly for a two-component mixture. Furthermore, we calculated these coefficients using integral
relations, derived in \cite{glav/tc1}. This gives an independent way to determine the interfacial resistances.

The main input parameters of the model are the local resistivity profiles used to calculate the continuous solution. There is not much theoretical information about
the numerical value of these resistivities. In the vapor phase one can use kinetic theory. In the liquid phase it is most appropriate to use experimental values.
There is no experimental information about the local resistivities in the interfacial region. As the local resistivities change in the surface from one bulk value
to the other, it is natural to assume that they contain a contribution similar to the profile of the order parameter. There is also evidence from molecular dynamics
simulations for one-component systems \cite{surfres} that there is a peak in the local resistivity in the surface. As we are in the framework of the gradient
theory, it is naturally to assume that such peaks are caused by a square gradient term, which is similar to the gradient contribution to the Helmholtz energy
density in the interfacial region, namely $|\nabla\rho|^{2}$. The amplitudes of these peaks are not given by any theory and were used as parameters. We therefore
get that the three local resistivities for a two-component mixture have the form given in \eqr{eq/Equations/Phenomenological/03}. Thus we get three adjustable
amplitudes, $\alpha_{qq}$, $\alpha_{1q}$ and $\alpha_{11}$, two of which were found to contribute significantly to the value of the transfer coefficients.

In order to determine the typical values of the $\alpha $'s we need to compare our results with independently obtained resistivities. Unfortunately, not much
experimental data are available for multi-component resistivities and, to the best of our knowledge, no data are available for our system. Furthermore, no molecular
dynamic simulations of these properties are available for mixtures. The only available source of comparison is kinetic theory, which gives the expressions for the
interfacial resistivities or transfer coefficients given in \eqr{eq/KineticTheory/01}. We therefore compare our results to kinetic theory. Having three adjustable
parameters in the gradient theory, $\alpha_{qq}$, $\alpha_{1q}$ and $\alpha_{11}$, and two adjustable parameters in kinetic theory, the condensation parameters
$\sigma_{1}$ and $\sigma_{2}$, we are able to match three diagonal coefficients $R_{qq}$, $R_{11}$ and $R_{22}$. We found that $R_{qq}$ does not really depend on
$\alpha_{1q}$ and $\alpha_{11}$. This makes it possible to fit $\alpha_{qq}$ using $R_{qq}$ alone. For the values of the temperature and chemical potentials
considered this gave $\alpha_{qq} \simeq 9$. We furthermore found that the interfacial resistivities did not really depend on $\alpha_{1q}$. We therefore took this
amplitude equal to zero. In kinetic theory $R_{11}$ and $R_{22}$ depend on the condensation coefficients $\sigma_{1}$ and $\sigma_{2}$, respectively. Choosing
$\alpha_{11}=3$ gives values for the condensation coefficients of 0.54 and 0.62. As the components considered are very similar it is to be expected that these
coefficients are close to each other. The values of $\alpha$'s obtained from the matching are such that the excess entropy production of the surface is positive,
the second law is obeyed and the Onsager relations are valid. Having found the values of the $\alpha$'s from the diagonal transfer coefficients the values of the
cross coefficients follow.

We found that the values of the cross coefficients, obtained by our method are an order of magnitude larger than those found from kinetic theory. This confirms
results from molecular dynamics simulations \cite{jialin/longrange} for a one-component system, where it was found that increasing the range of the attractive
potential increased in particular the cross coefficients substantially above the values predicted by kinetic theory. This is an interesting result, indicating that
kinetic theory underestimates the transfer coefficients for real fluids. This also indicates, that the effect of coupling will be important in the interfacial
region. Experiments also confirm the importance of the cross coefficients \cite{Mills2002, James2006}.

The effect of cross coefficients can be related to the measurable quantities, such as measurable heat of transfer $q^{*}_{i} \equiv - R_{qi}/R_{qq}$. This quantity
can be associated both with gas and liquid phases in accordance to the corresponding heat fluxes. The difference $q^{*,g}_{i} - q^{*,\ell}_{i} = -
(R_{qi}^{g}-R_{qi}^{\ell})/R_{qq} = - (h_{i,eq}^{g}-h_{i,eq}^{\ell})$ is equal to the difference of partial enthalpies between gas and liquid in
equilibrium\footnote{We note that $R_{qq}^{g} = R_{qq}^{\ell} \equiv R_{qq}$.}. This quantity is substantial, which implies that $q^{*,g}_{i} - q^{*,\ell}_{i}$ is
also substantial. This, in turn, makes the difference between the cross coefficients on the vapor and the liquid side substantial. This gives a theoretical ground
for the importance of coupling in the interfacial region. Experiments \cite{Mills2002, James2006} confirm the size and importance of the heat of transfer on the
vapor side.

We did the comparison for one value of the temperature and chemical potential only. If one extends the analysis to a larger domain, one finds that the $\alpha$'s
depend on the temperature and the chemical potential difference; we refer to \cite{bedeaux/vdW/III} in this context. The results of kinetic theory \cite{Pao1971a,
Sone1973, Cipolla1974, Bedeaux1990} and molecular dynamics \cite{Simon2004} both support the existence of a peak in the diagonal local resistivities and therefore
the use of finite values for $\alpha_{qq}$ and $\alpha_{11}$.

Furthermore, it was found, that the data obtained directly from non-equilibrium numerical solution agree with the ones obtained using integral relations, as is
expected. This gives an alternative and easier way to determine non-equilibrium properties of the interfacial region, needing only equilibrium information about the
system. In fact, as we speak of linear non-equilibrium thermodynamics, this is the way it should be. The interfacial resistances are determined from equilibrium
properties, just like Green-Kubo relations involve only equilibrium information in order to determine the transport coefficients.

\appendix

\section{Determining an optimal perturbation rate}\label{sec/Perturbation}

The value of the resistance $\mathrm{R}(T_{eq}, \mu_{12,\,eq})$ does not depend on the perturbation, given the perturbation is small enough. However, the magnitude
$\ss$ of the perturbation which may be considered sufficiently small, has to be determined empirically. This would require considering perturbations where $\ss$ is
beyond the appropriate range and will make the empirical resistances $\mathrm{R}(T_{eq}, \mu_{12,\,eq})$ to be dependent on $\ss$.

In order to determine the appropriate range of perturbations, that is when $\ss$ is small enough to consider them linear, and at the same time, large enough, to not
interfere with the accuracy of the numerical solution, we check the obtained resistances for the thermodynamic consistency. We have the following constraints, which
they must obey for each $T$ and $\mu_{12}$:

- i) the cross coefficients of each $\mathrm{R}$ matrix must satisfy Onsager relations;

- ii) the second law consistency;

- iii) coefficients obtained on the gas and the liquid side of the surface must be related;

We will use the first condition to determine the range of $\ss$, while the two remaining will be used for the verification of the results obtained in the paper.

\subsection{Onsager reciprocal relations}

As shown by Onsager \cite{onsager/rec}, the cross coefficients must be the same. We therefore have $R_{qi} = R_{iq}$ and $R_{ji} = R_{ij}$.

We calculate the coefficients at the values of equilibrium temperature and chemical potential difference $T_{eq} = 330$ K and $\mu_{12,\,eq} = 700$ J/mol for
different values of the adjustable amplitudes $\alpha_{qq}$, $\alpha_{1q}$, and $\alpha_{11}$.

In \tblsr{tbl/Onsager/Pc-beta-0-0-0}{tbl/Onsager/Ex-beta-0-0-0} we give the relative error in percent for the gas-side cross coefficients
$|(R_{ij}^{g}-R_{ji}^{g})/R_{ij}^{g}|\spd 100\%$ as a function of $\beta $ for $\alpha_{qq}=0$, $\alpha_{1q}=0$, $\alpha_{11}=0$ obtained by different methods.
\begin{longtable}{l@{\qquad}l@{\qquad}l@{\qquad}l@{\qquad}l@{\qquad}l@{\qquad}l}%
\caption{Relative error in percent for gas-side cross-coefficients obtained by "perturbation cell" method at $T_{eq} = 330$ and $\mu_{12,\,eq} = 700$ for different $\beta$ and for $\alpha_{qq} = 0$, $\alpha_{1q} = 0$, $\alpha_{11} = 0$.} \label{tbl/Onsager/Pc-beta-0-0-0}\\%
\hline %
$\beta$  & $R_{q1}$  & $R_{q2}$  & $R_{12}$  \\%
\hline %
2.0e-002     & 8.963066 & 35.863259 & 34.908631 \\%
2.0e-003     & 0.273286 & 0.369082  & 19.683274 \\%
2.0e-004     & 0.011726 & 0.007231  & 1.909391  \\%
2.0e-005     & 0.066375 & 0.071266  & 2.336652  \\%
2.0e-006     & 4.963895 & 8.128243  & 5.843913  \\%
\hline %
\end{longtable} %
\begin{longtable}{l@{\qquad}l@{\qquad}l@{\qquad}l@{\qquad}l@{\qquad}l@{\qquad}l}%
\caption{Relative error in percent for gas-side cross-coefficients obtained by "experiment like" method at $T_{eq} = 330$ and $\mu_{12,\,eq} = 700$ for different $\beta$ and for $\alpha_{qq} = 0$, $\alpha_{1q} = 0$, $\alpha_{11} = 0$.} \label{tbl/Onsager/Ex-beta-0-0-0}\\%
\hline %
$\beta$  & $R_{q1}$  & $R_{q2}$  & $R_{12}$  \\%
\hline %
2.0e-002     & 1.275105 & 0.828600  & 754.982200    \\%
2.0e-003     & 0.038759 & 0.363715  & 38.708981 \\%
2.0e-004     & 0.131868 & 0.238584  & 6.247648  \\%
2.0e-005     & 1.301483 & 2.056102  & 20.984734 \\%
2.0e-006     & 13.282959    & 20.788752 & 632.124504    \\%
\hline %
\end{longtable} %
As one can see, $\beta =0.02$ is really an extreme perturbation and the difference is rather large. When we decrease $\beta $ to 2e-4 the  differences become small.
As we further decrease $\beta$ to 2e-6 the inaccuracy of the numerical solution become comparable to the size of the perturbation. We conclude that the values for
$\beta$ to 2e-4 are closest to the converged values and use them as such.

In \tblsr{tbl/Onsager/Pc-beta-10-10-10}{tbl/Onsager/Ex-beta-10-10-10} we give the same data for the higher continuous resistivities with rather  substantial peak,
when $\alpha_{qq} = 10$, $\alpha_{1q} = 10$ and $\alpha_{11} = 10$. As one can see, the Onsager relations are fulfilled there again best for $\beta=$ 2e-4
\begin{longtable}{l@{\qquad}l@{\qquad}l@{\qquad}l@{\qquad}l@{\qquad}l@{\qquad}l}%
\caption{Relative error in percent for gas-side cross-coefficients obtained by "perturbation cell" method at $T_{eq} = 330$ and $\mu_{12,\,eq} = 700$ for different $\beta$ and for $\alpha_{qq} = 10$, $\alpha_{1q} = 10$, $\alpha_{11} = 10$.} \label{tbl/Onsager/Pc-beta-10-10-10}\\%
\hline %
$\beta$  & $R_{q1}$  & $R_{q2}$  & $R_{12}$  \\%
\hline %
2.0e-002     & 71.515410    & 78.166809 & 23.572836 \\%
2.0e-003     & 0.745604 & 0.896547  & 0.317348  \\%
2.0e-004     & 0.012358 & 0.012650  & 0.001919  \\%
2.0e-005     & 0.012078 & 0.007485  & 0.005290  \\%
2.0e-006     & 0.713969 & 1.124994  & 0.022121  \\%
\hline %
\end{longtable} %
\begin{longtable}{l@{\qquad}l@{\qquad}l@{\qquad}l@{\qquad}l@{\qquad}l@{\qquad}l}%
\caption{Relative error in percent for gas-side cross-coefficients obtained by "experiment like" method at $T_{eq} = 330$ and $\mu_{12,\,eq} = 700$ for different $\beta$ and for $\alpha_{qq} = 10$, $\alpha_{1q} = 10$, $\alpha_{11} = 10$.} \label{tbl/Onsager/Ex-beta-10-10-10}\\%
\hline %
$\beta$  & $R_{q1}$  & $R_{q2}$  & $R_{12}$  \\%
\hline %
2.0e-002     & 4.225362 & 2.559393  & 12.259260 \\%
2.0e-003     & 0.443944 & 0.256804  & 1.091842  \\%
2.0e-004     & 0.068621 & 0.019788  & 0.093041  \\%
2.0e-005     & 0.269764 & 0.407090  & 0.008844  \\%
2.0e-006     & 2.717575 & 4.149484  & 2.025054  \\%
\hline %
\end{longtable} %

We may notice that the behavior of the resistivities with respect to $\beta$ is independent on the behavior of the resistivities with respect to $\alpha_{qq}$,
$\alpha_{1q}$ and $\alpha_{11}$. This is natural, as these parameters control the different aspects of the system: $\beta$ controls the perturbation rate, while
$\alpha$'s are adjustable parameters, which control the size of the peak in the continuous resistivities.

\section{Consistency of the non-equilibrium solution}\label{sec/Consistency}
\subsection{Second law consistency}

In this subsection we investigate the values of parameters $\alpha_{qq}$, $\alpha_{1q}$, $\alpha_{11}$ for which the second law of thermodynamics is fulfilled. That
is that the excess entropy production is positive and therefore the matrix of the resistivity coefficients is positive definite. This requires that the diagonal
coefficients  are positive and for each pair $q1$, $q2$ and $12$ of the cross coefficients the expression
\begin{equation}\label{eq/Results/01}
DR_{ik} \equiv R_{ii}R_{kk} - {1 \over 4}(R_{ik}+R_{ki})^{2} > 0
\end{equation}
must be positive.

In \tblr{tbl/2ndLaw/Pc-4-Aqq-0-0} we give the diagonal coefficients and expression \eqref{eq/Results/01} for each pair of the cross coefficients as a function of
$\alpha_{qq}$ for $\alpha_{1q} = 0$, $\alpha_{11} = 0$ and $\beta =$ 2e-4 obtained by the "perturbation cell" method. In
\tblsr{tbl/2ndLaw/Pc-4-0-Aiq-0}{tbl/2ndLaw/Pc-4-0-0-Aii} we give the same quantities for other choices of $\alpha$.
\begin{longtable}{l@{\qquad}l@{\qquad}l@{\qquad}l@{\qquad}l@{\qquad}l@{\qquad}l}%
\caption{2nd law consistency for gas-side coefficients. The diagonal coefficients and the quantities defined by \eqref{eq/Results/01}. Data are obtained by "perturbation cell" method at $T_{eq} = 330$ and $\mu_{12,\,eq} = 700$ for different $\alpha_{qq}$ and for $\beta = 0.0002$, $\alpha_{1q} = 0$, $\alpha_{11} = 0$.} \label{tbl/2ndLaw/Pc-4-Aqq-0-0}\\%
\hline %
$\alpha_{qq}$    & $R_{qq}$  & $R_{11}$  & $R_{22}$  & $DR_{q1}$  & $DR_{q2}$  & $DR_{12}$  \\%
\hline %
0    & 7.05644e-015 & 0.0754717 & -0.0919278    & 2.13025e-015  & -2.59473e-015 & -0.0277518    \\%
1    & 3.36047e-012 & 0.0937784 & -0.0741586    & 1.26056e-012  & -9.9683e-013  & -0.0278179    \\%
10   & 3.35408e-011 & 0.259425  & 0.0851534 & 3.48053e-011  & 1.14244e-011  & 0.0874467 \\%
\hline %
\end{longtable} %
\begin{longtable}{l@{\qquad}l@{\qquad}l@{\qquad}l@{\qquad}l@{\qquad}l@{\qquad}l}%
\caption{2nd law consistency for gas-side coefficients. The diagonal coefficients and the quantities defined by \eqref{eq/Results/01}. Data are obtained by "perturbation cell" method at $T_{eq} = 330$ and $\mu_{12,\,eq} = 700$ for different $\alpha_{1q}$ and for $\beta = 0.0002$, $\alpha_{qq} = 0$, $\alpha_{11} = 0$.} \label{tbl/2ndLaw/Pc-4-0-Aiq-0}\\%
\hline %
$\alpha_{1q}$    & $R_{qq}$  & $R_{11}$  & $R_{22}$  & $DR_{q1}$  & $DR_{q2}$  & $DR_{12}$  \\%
\hline %
0    & 7.05644e-015 & 0.0754717 & -0.0919278    & 2.13025e-015  & -2.59473e-015 & -0.0277518    \\%
1    & 7.05608e-015 & 0.0746391 & -0.0910331    & 2.10664e-015  & -2.56935e-015 & -0.0271785    \\%
10   & 7.05304e-015 & 0.0670813 & -0.0828915    & 1.89251e-015  & -2.33855e-015 & -0.0222419    \\%
\hline %
\end{longtable} %
\begin{longtable}{l@{\qquad}l@{\qquad}l@{\qquad}l@{\qquad}l@{\qquad}l@{\qquad}l}%
\caption{2nd law consistency for gas-side coefficients. The diagonal coefficients and the quantities defined by \eqref{eq/Results/01}. Data are obtained by "perturbation cell" method at $T_{eq} = 330$ and $\mu_{12,\,eq} = 700$ for different $\alpha_{11}$ and for $\beta = 0.0002$, $\alpha_{qq} = 0$, $\alpha_{1q} = 0$.} \label{tbl/2ndLaw/Pc-4-0-0-Aii}\\%
\hline %
$\alpha_{11}$    & $R_{qq}$  & $R_{11}$  & $R_{22}$  & $DR_{q1}$  & $DR_{q2}$  & $DR_{12}$  \\%
\hline %
0    & 7.05644e-015 & 0.0754717 & -0.0919278    & 2.13025e-015  & -2.59473e-015 & -0.0277518    \\%
1    & 7.05717e-015 & 0.370078  & 0.265626  & 1.04468e-014  & 7.49827e-015  & 0.381226  \\%
10   & 7.10378e-015 & 3.02063   & 3.48284   & 8.58316e-014  & 9.89654e-014  & -69.2846  \\%
\hline %
\end{longtable} %

We see, that the required quantities become positive for rather big values of $\alpha_{qq}$. They almost do not depend on the value of  $\alpha_{1q}$ and they are
positive for moderate values of parameter $\alpha_{11}$. It is clear that finite values of $\alpha_{qq}$ and $\alpha_{11}$ are needed to have a positive excess
entropy production.

All the above quantities almost do not depend on the value of $\beta$ in the range [1e-5, 1e-3]. The "experimental-like" procedure leads to almost the same values
of all the quantities. The liquid-side coefficients reveal a similar behavior.

\subsection{Gas- and liquid- coefficients}

One can use the measurable flux $J_{q}'$ extrapolated from the liquid side of the surface, rather then from the gas side, using \eqr{eq/ExcessEntropy/03}. In this
case one should use not the enthalpy of the gas bulk $h^{g}$, but enthalpy of the liquid bulk $h^{\ell}$ extrapolated to the diving surface. The two measurable
fluxes are related as
\begin{equation}\label{eq/TransportCoef/03}
J_{q}^{\,\prime,\,g} - J_{q}^{\,\prime,\,\ell} = \sum_{i=1}^{n}{J_{\xi_i}\left(\tilde{h}_{i,\,eq}^{\ell}-\tilde{h}_{i,\,eq}^{g}\right)}
\end{equation}%
Identifying the forces and fluxes and writing the linear force-flux relations for the measurable heat flux on the liquid side, one introduces the interfacial
resistances measured on the liquid side in the same way as it was done in \secr{sec/EntropyProduction} for the interfacial resistances measured on the gas side.
These resistances are related as follows
\begin{equation}\label{eq/TransportCoef/04}
\begin{array}{rl}
R_{qq}^{\ell} &= R_{qq}^{g}\\\\
R_{qi}^{\ell} + {h}_{i,\,eq}^{\ell}\,R_{qq}^{\ell} &= R_{qi}^{g} + {h}_{i,\,eq}^{g}\,R_{qq}^{g} \\\\
R_{iq}^{\ell} + {h}_{i,\,eq}^{\ell}\,R_{qq}^{\ell} &= R_{iq}^{g} + {h}_{i,\,eq}^{g}\,R_{qq}^{g} \\\\
R_{ji}^{\ell} + {h}_{i,\,eq}^{\ell}\,R_{jq}^{\ell} + {h}_{j,\,eq}^{\ell}\,R_{qi}^{\ell} + {h}_{i,\,eq}^{\ell}\,{h}_{j,\,eq}^{\ell}\,R_{qq}^{\ell} &= %
R_{ji}^{g} + {h}_{i,\,eq}^{g}\,R_{jq}^{g} + {h}_{j,\,eq}^{g}\,R_{qi}^{g} + {h}_{i,\,eq}^{g}\,{h}_{j,\,eq}^{g}\,R_{qq}^{g}\\\\
\end{array}
\end{equation}
These coefficients can be calculated independently from a non-equilibrium numerical solution. Given that, the validity of \eqr{eq/TransportCoef/04} would indicate
the internal consistency of the model. In this subsection we verify these relations.

In \tblr{tbl/Bulk/Pc-4-1-1-1} we give the relative error in  percent between the left hand side and the right hand side of \eqr{eq/TransportCoef/04}.
\begin{longtable}{l@{\qquad}l@{\qquad}l@{\qquad}l@{\qquad}l@{\qquad}l@{\qquad}l@{\qquad}l@{\qquad}l@{\qquad}l}%
\caption{Relative error in percent for invariant expressions in \eqr{eq/TransportCoef/04} obtained by "perturbation cell" method at $T_{eq} = 330$ and $\mu_{12,\,eq} = 700$ for $\beta = 0.0002$ and $\alpha_{qq} = 1$, $\alpha_{1q} = 1$, $\alpha_{11} = 1$.} \label{tbl/Bulk/Pc-4-1-1-1}\\%
\hline %
${qq}$     & ${11}$  & ${22}$  & ${q1}$  & ${1q}$  & ${q2}$  & ${2q}$  & ${12}$  & ${21}$   \\%
\hline %
 0.000000   & 0.000002  & 0.000085  & 0.000001  & 0.000389  & 0.000001  & 0.000389  & 0.000060  & 0.000003  \\%
\hline %
\end{longtable} %
For instance, the ${q1}$ quantity is equal to $|(R_{q1}^{\ell} - {h}_{1,\,eq}^{\ell}\,R_{qq}^{\ell}) - (R_{q1}^{g} - {h}_{1,\,eq}^{g}\,R_{qq}^{g})|/|R_{q1}^{\ell} -
{h}_{1,\,eq}^{\ell}\,R_{qq}^{\ell}|\spd100\%$. The other quantities are defined in the same way. These errors almost do not depend neither  on the value of $\beta$
in the range [1e-5, 1e-3] nor on the values of $\alpha_{qq}$, $\alpha_{1q}$, $\alpha_{11}$. The "experimental-like" procedure leads to almost the same results.

\subsection{Integral relations}

For two component mixture the force-flux equations have a form
\begin{equation}\label{eq/Results/01}
\begin{array}{rl}
X_{q} &= R^{\,\prime}_{qq}\,J_{q}^{\,\prime} -  R^{\,\prime}_{q1}\,J_{\xi_{1}} - R^{\,\prime}_{q2}\,J_{\xi_{2}} \\
X_{1} &= R^{\,\prime}_{1q}\,J_{q}^{\,\prime} -  R^{\,\prime}_{11}\,J_{\xi_{1}} - R^{\,\prime}_{12}\,J_{\xi_{2}} \\
X_{2} &= R^{\,\prime}_{2q}\,J_{q}^{\,\prime} -  R^{\,\prime}_{21}\,J_{\xi_{1}} - R^{\,\prime}_{22}\,J_{\xi_{2}} \\
\end{array}
\end{equation}%
The left hand side of each equation must be equal to the right hand side. The difference therefore reflects the error. We give the relative error between the left
an the right hand side of \eqr{eq/Results/01} in percent in \tblr{tbl/X/Pc-Ir-4-9-0-3}. As a testing perturbation we used one of those used in the perturbation cell
method.
\begin{longtable}{l@{\qquad}l@{\qquad}l@{\qquad}l@{\qquad}l@{\qquad}l@{\qquad}l@{\qquad}l@{\qquad}l@{\qquad}l}%
\caption{Relative error in percent between the left- and right- hand side of \eqr{eq/Results/01} for coefficients obtained by "perturbation cell" and "integral relations" methods at $T_{eq} = 330$ and $\mu_{12,\,eq} = 700$ for $\beta = 0.0002$ and $\alpha_{qq} = 9$, $\alpha_{1q} = 0$, $\alpha_{11} = 3$ .} \label{tbl/X/Pc-Ir-4-9-0-3}\\%
\hline
                & \multicolumn{3}{c}{Integral relations}    & \multicolumn{3}{c}{Perturbation cell} \\
\hline %
phase & $X_{q}$     & $X_{1}$  & $X_{2}$  & $X_{q}$  & $X_{1}$  & $X_{2}$ \\%
\hline %
gas     & 0.059489    & 0.037918  & 0.296959  & 0.046965    & 0.087411  & 0.867098  \\%
liquid  & 0.059489    & 0.172608  & 0.027275  & 0.046851    & 0.216819  & 0.014248  \\%
\hline %
\end{longtable} %
Again, the relative difference is not more then a few promille. Given that this is the case even for a few percent difference in one of the coefficients, we may
conclude that the values of the forces are insensitive to the precise value of this resistivity coefficient. This also indicates that the value of this coefficient
obtained in \cite{glav/grad3} has a 6\% error. This does not necessarily affect, however, the accuracy of the integral relations.

\bibliographystyle{unsrt}

\end{document}